# Crucial role of subsurface ocean variability in tropical cyclone genesis

Cong Gao [1,12], Lei Zhou [1,2] ✉, I.-I. Lin [3], Chunzai Wang [4,5,6], Shoude Guan [7,8], Fei-Fei Jin [9] & Raghu Murtugudde [10,11]

The upper ocean provides thermal energy to tropical cyclones. However, the impacts of the subsurface ocean on tropical cyclogenesis have been largely overlooked. Here, we show that the subsurface variabilities associated with the variation in the 26 °C isothermal depth have pronounced impacts on tropical cyclogenesis over global oceans. The sea surface wind stress and its curl before tropical cyclogenesis are large enough to perturb the ocean interior down to more than one hundred meters due to entrainment and upwelling. The 26 °C isothermal depth can fluctuate by tens of meters to significantly modify the upper ocean heat content. Consequently, sea surface temperature anomalies under nascent tropical cyclones are induced, and tropical cyclogenesis is modulated. Our results substantiate an unexpected relation between ocean interior variations and tropical cyclogenesis.

Tropical cyclones (TCs), producing damaging winds[1], heavy rainfalls[2], storm surges[3], and inland floods[4], are among the deadliest natural disasters worldwide[5–7]. During 1970–2019, TCs resulted in 779,324 fatalities and 1.4 trillion U.S. dollars in economic losses[8]. As coastal cities continue to become more populated and developed, the need for TC risk assessment and management becomes increasingly critical[4,9]. The locations and frequencies of TC genesis modulate the lifetime maximum intensities of TCs and exposure to hazards[10–12]. Therefore, enhancing our understanding of TC genesis is crucial for reducing TC-related risks and devastation[13,14], which are expected to worsen with continued global warming.

TC genesis is determined by many environmental properties, such as the earth's rotational effects, the atmospheric parameters of low-level relative vorticity[15], vertical wind shear[16], mid-level relative humidity[17], and atmospheric instability[18]. Ocean is the energy source for TCs[19,20]. The upper ocean heat content[21] is also essential for TC genesis, which was recognized long ago[22]. Based on observations, the "ocean thermal energy", representing the ocean heat content above the depth of the 26 °C isotherm (D26; Fig. 1a), was adopted as one of the six parameters in the first empirical model for TC genesis (i.e., the TC genesis potential index established in ref. 22). However, there has been disagreement on the influence of the upper ocean heat content on TC genesis, and numerical simulations have drawn different conclusions[23]. The controversy has grown because satellite-observed sea surface temperatures (SSTs) do not capture the subsurface conditions. In contrast, the impacts of upper ocean heat content (or generally variabilities in the subsurface ocean) on TC intensity, rather than on TC genesis, have been well examined[24–27]. SST was replaced with the inclusion of subsurface temperature information in the potential intensity index, yielding a better representation of TC intensity and improving its forecast[28]. Nevertheless, SSTs are assumed to play a dominant role in TC

[1]School of Oceanography, Shanghai Jiao Tong University, Shanghai, China. [2]Southern Marine Science and Engineering Guangdong Laboratory (Zhuhai), Zhuhai, China. [3]Department of Atmospheric Sciences, National Taiwan University, Taipei, Taiwan. [4]State Key Laboratory of Tropical Oceanography, South China Sea Institute of Oceanology, Chinese Academy of Sciences, Guangzhou, China. [5]Global Ocean and Climate Research Center, South China Sea Institute of Oceanology, Chinese Academy of Sciences, Guangzhou, China. [6]Guangdong Key Laboratory of Ocean Remote Sensing, South China Sea Institute of Oceanology, Chinese Academy of Sciences, Guangzhou, China. [7]Frontier Science Centre for Deep Ocean Multispheres and Earth System (FDOMES), Ocean University of China, Qingdao, China. [8]Physical Oceanography Laboratory, Ocean University of China, Qingdao, China. [9]Department of Atmospheric Sciences, University of Hawaii, Honolulu, HI, USA. [10]Department of Atmospheric and Oceanic Science, University of Maryland, College Park, MD, USA. [11]Inter-Disciplinary Program in Climate Studies, Indian Institute of Technology Bombay, Mumbai, India. [12]Present address: Department of Civil and Environmental Engineering, Princeton University, Princeton, NJ, USA. ✉e-mail: zhoulei1588@sjtu.edu.cn

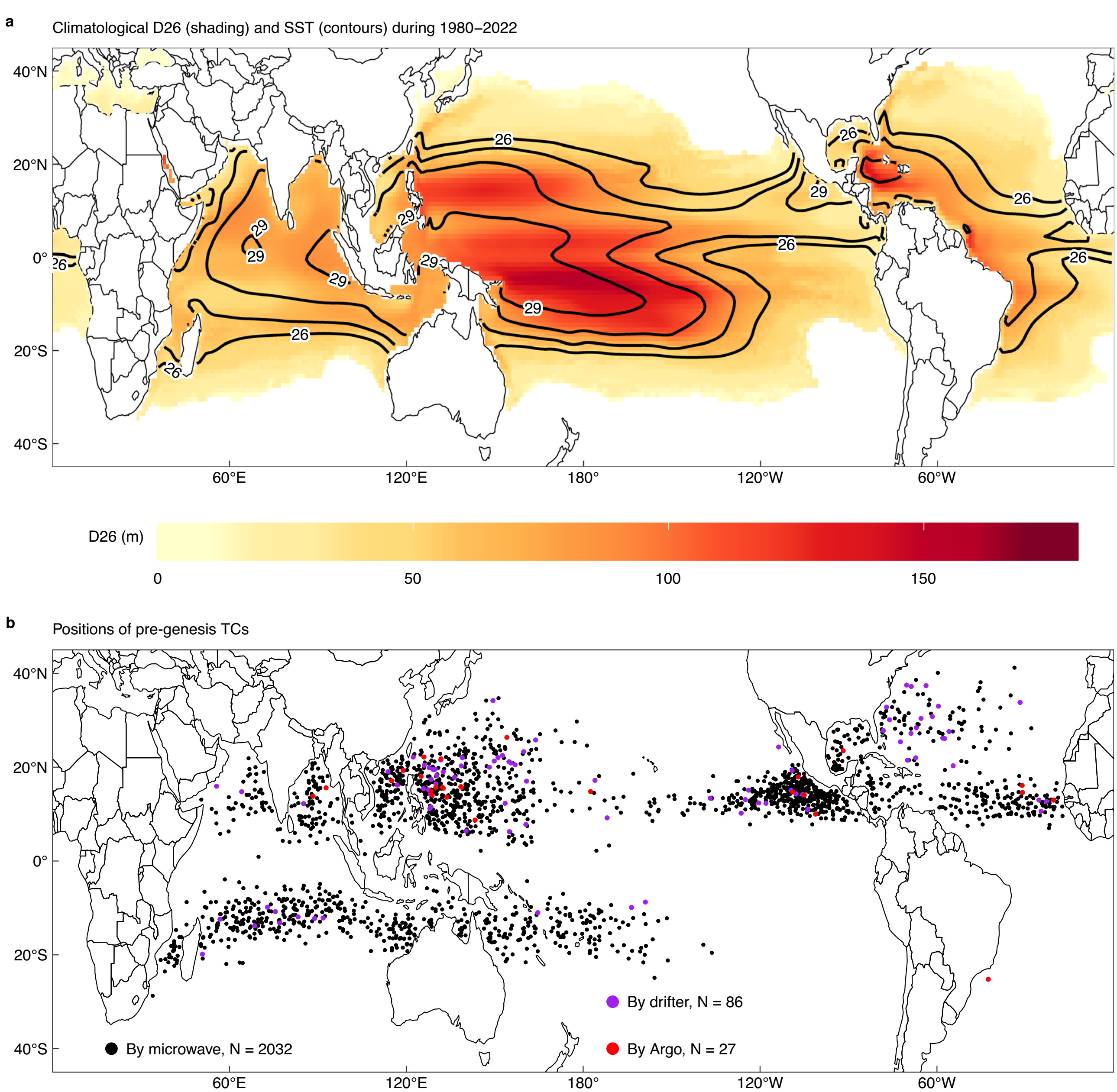

**Fig. 1 | Climatological sea surface and subsurface environments and positions of pre-genesis tropical cyclones (TCs). a** Climatological depth of the 26 °C isotherm (D26) from 1980 to 2022 (colour shading). The black contours denote the climatological sea surface temperatures (SSTs) from 26 °C to 29 °C. **b** Positions of pre-genesis TCs with observed SSTs. The positions where SSTs are observed by microwave satellites are denoted with black dots. The positions observed by the drifters are denoted with purple dots, and those observed by the Argo floats are denoted with red dots.

genesis in most studies[13,19,23,29], which is different from the original idea[22].

The reason that the upper ocean heat content was largely overlooked in TC genesis may be the fallacious assumption that the "weak" winds during TC genesis could not induce pronounced variabilities in the ocean interior. Consequently, the likelihood of the subsurface ocean influence on TC genesis has generally been dismissed. Practically, the depth influenced by TC winds can serve as a proxy for the depth of cold water that is drawn up into the upper mixed layer and has impacts on ocean stratification, D26, and upper ocean heat content. In fact, in-situ subsurface mooring observations demonstrated that the immediate influence of Hurricane Frederic (1979) could extend from the surface down to as deep as 950 m[30,31], underscoring the importance of TC-subsurface ocean interactions. A moderate TC with a maximum wind speed of 20–40 m s$^{-1}$ deepened the ocean mixed layer by 90 m and led to a strong Ekman upwelling velocity of 2 × 10$^{-3}$ m s$^{-1}$ (which was equivalent to a vertical displacement of 86 m in half a day)[32]. A recent study revealed the negative role of the subsurface ocean in TC genesis due to Ekman upwelling during El Niño, contrasting with the positive role of SSTs[33].

In this study, we show that variabilities in the subsurface ocean have detectable impacts on TC genesis by examining the genesis stages of 2032 global TCs over the past 25 years from 1998 to 2022. By definition[34], the maximum wind speed is less than 35 knots (18 m s$^{-1}$; 1 knot ≈ 0.51 m s$^{-1}$) during the pre-genesis stage. We show that this wind is strong enough to displace D26 and thus modify the upper ocean heat content. Consequently, SST variabilities due to TC's entrainment mixing and the Ekman effect by wind stress curl beneath nascent TCs are prominent enough to influence TC genesis. Our findings highlight the role of the subsurface ocean in TC genesis. Considering the out-of-sync warming trends in the ocean surface and subsurface under climate change[27], it is essential to explore the impacts of the subsurface ocean on TC genesis and their evolution under future scenarios. These findings also demand additional attention to model accuracy in winds, coupled air-sea interactions, and particularly ocean model initialization to better represent variabilities in D26 and ocean heat content in a warming world.

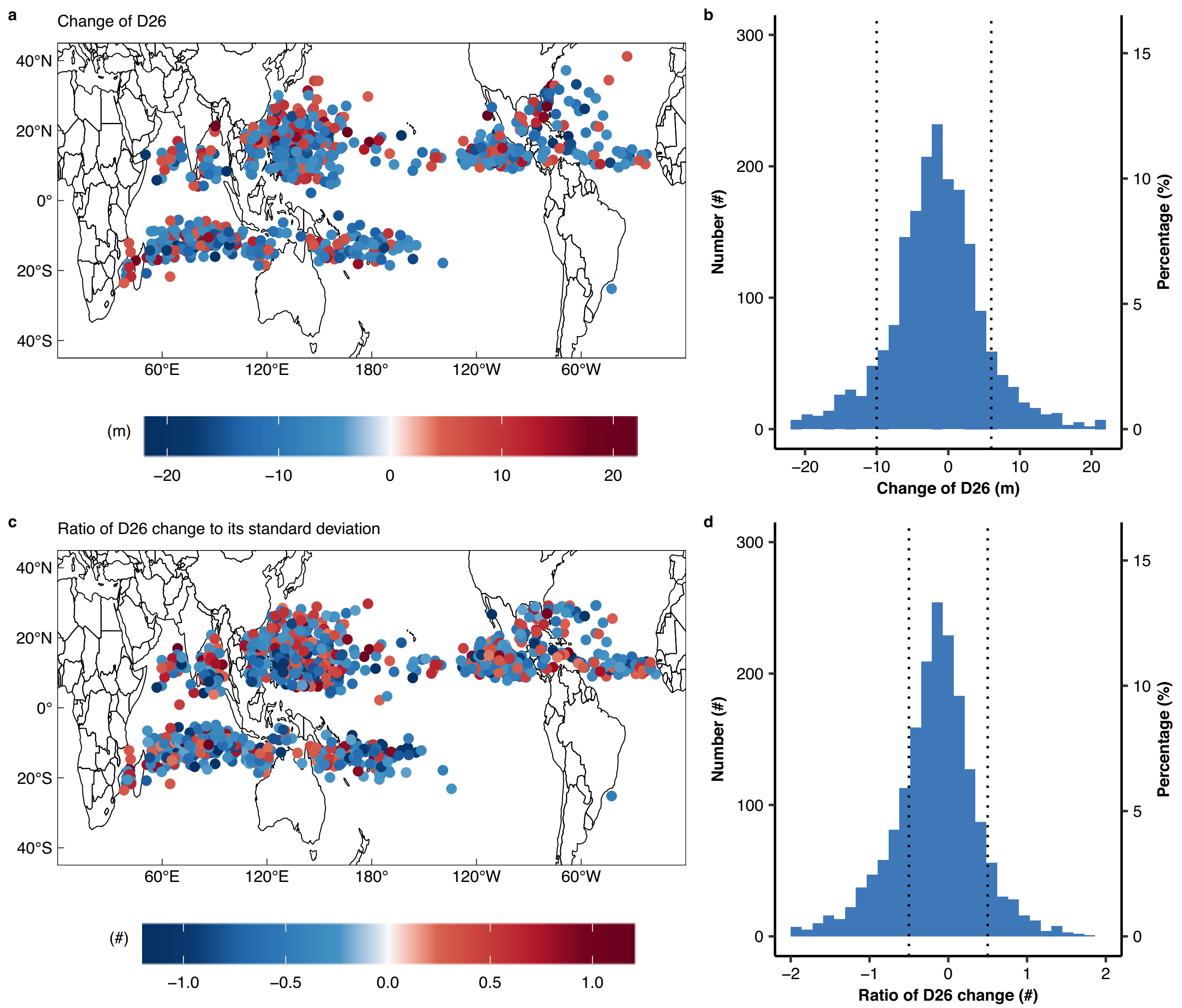

**Fig. 2 | Responses of subsurface ocean to pre-genesis tropical cyclones (TCs).** **a** Changes in depth of the 26 °C isotherm (D26) due to pre-genesis TCs. **b** Histogram of D26 changes. The dotted lines represent the 10th and 90th percentiles, respectively. **c**, Same as **a** but for the ratios of D26 change to the standard deviation of D26. **d** Same as **b** but for the ratios in **c**. The dotted lines represent ratios of −0.5 and 0.5, respectively. The D26 changes are obtained from the daily GLORYS12v1 product. The standard deviation of D26 is obtained by deriving it from the monthly GLORYS12v1 product and applying a 1-year high-pass filter to remove interannual and decadal variabilities. For each grid, the standard deviation is calculated using the monthly D26 data from 1993 to 2022.

## Significant disturbances in the ocean interior by pre-genesis TCs

According to the World Meteorological Organization, a TC is born when its intensity reaches 35 knots (18 m $s^{-1}$) or, equivalently, T2.5 according to the Dvorak technique[35]. A TC is considered in the pre-genesis stage (see Methods) when its intensity reaches 30 knots or T2.0 in the Dvorak technique. The first day of the pre-genesis stage is referred to as Day 0 and the negative (positive) days represent the days before (after) Day 0 hereafter. The locations and intensities of TCs are obtained from the best-track records (Methods). A total of 2032 pre-genesis TCs are identified from 1998 to 2022 (Fig. 1b).

Temperature profiles are obtained from the Global Ocean Reanalysis and Simulations (GLORYS) 12v1 product[36]. D26 is used to quantify subsurface variabilities, specifically, the ocean heat content in the upper layer[7,33,37,38]. The differences between the median D26 from Days −10 to −4 and D26 on Day 0 (denoted as ΔD26 hereafter; Methods) are shown in Fig. 2a. The positive ΔD26 is attributable to the vertical mixing (or so-called entrainment)[39,40], which cools the surface water and warms the subsurface water. A negative ΔD26 is induced by Ekman upwelling[39,41], which cools both the surface and the subsurface water. It is also consistent with the results of ref. 42 that D26 can be both shallowed by TC-induced upwelling and deepened by TC-induced downwelling, which combines the effect of vertical mixing. A maximum D26 deepening of 53 m and a maximum D26 shoaling of 48 m are observed (Fig. 2a). The pre-genesis winds of 10% of the TCs induce at least a 6-m increase or 10-m decrease in D26 (Fig. 2b). The ratios of ΔD26 to the local standard deviation of the monthly D26 variabilities are shown in Fig. 2c. Thirty-two percent of pre-genesis TCs perturb D26 by more than 50% of monthly D26 variabilities (Fig. 2d). Specifically, 10% pre-genesis TCs-induced ΔD26 is larger than +50%, and 22% pre-genesis TCs-induced ΔD26 is smaller than −50%.

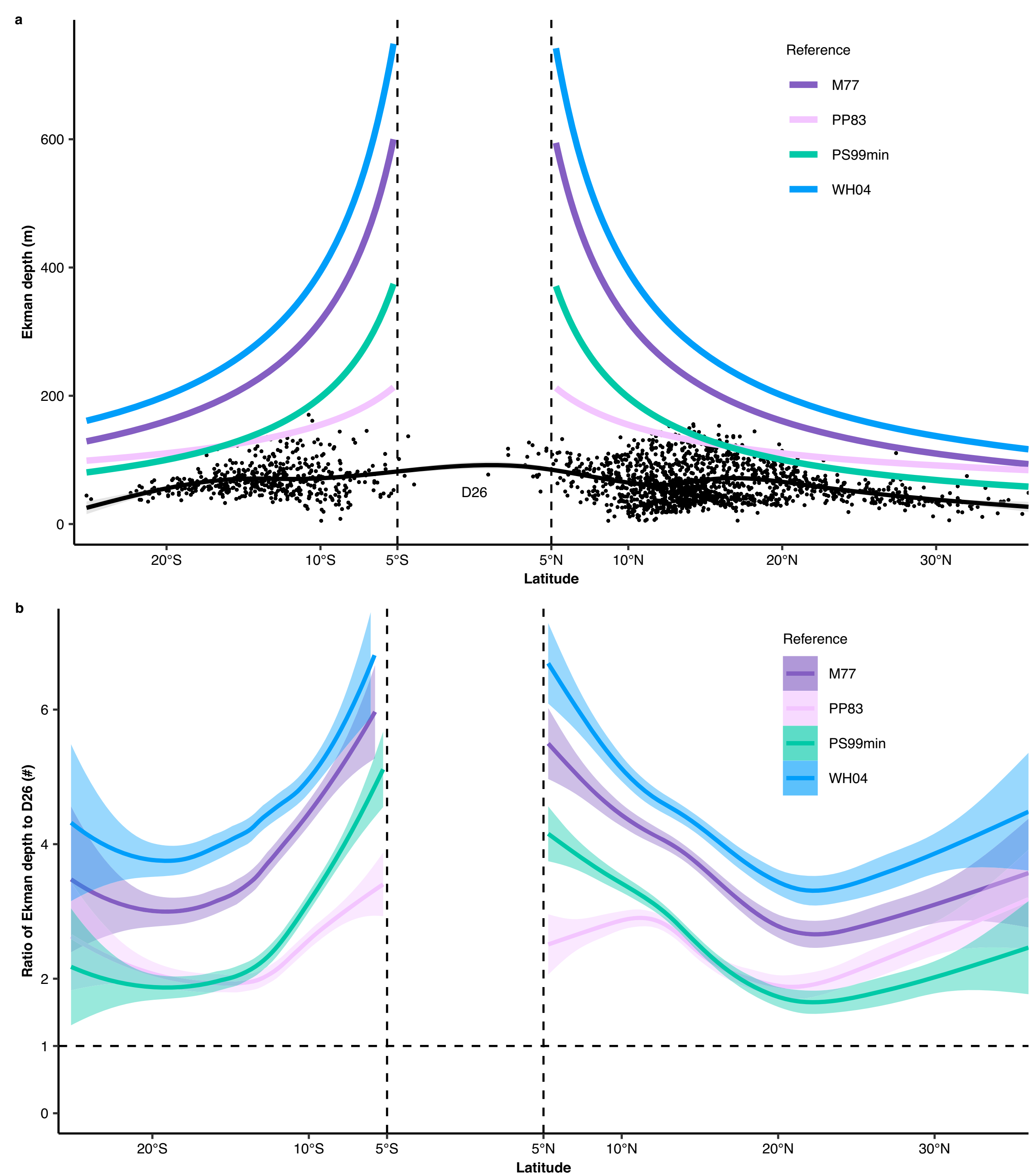

**Fig. 3 | Comparisons between Ekman layer depth and depth of the 26 °C isotherm (D26). a**, Coloured thick lines are the Ekman layer depths estimated with various empirical formulas (see Methods). Each black dot represents the D26 under a pre-genesis tropical cyclone (TC) case. The black line is the mean D26 of all pre-genesis TCs. **b** Ratios of the estimated Ekman layer depth to the D26 under pre-genesis TCs. The shaded areas represent the 95% confidence intervals.

## Dynamics of the deep penetration of wind forcing during pre-genesis of TCs

The wind speeds during the pre-genesis stage of a TC are much lower than those during its mature stage. Thus, it has traditionally been presumed that weak winds during pre-genesis cannot have significant impacts on the subsurface ocean. However, in nature, the wind speeds are already greater than 15 m $s^{-1}$ during TC pre-genesis, which can be as much as Force 8 on the Beaufort scale. The associated sea state codes are usually 6–7, which means a probable wave height of approximately 6 m, and the maximum wave height can reach 7.5 m. Thus, the winds during TC pre-genesis are not weak at all. In addition, the wind stress curl is very efficient in driving the ocean via Ekman pumping and suction[43–46]. The climatological near-surface wind stress curl is on the order of $10^{-7}$ kg $m^{-2}$ $s^{-2}$ (Supplementary Fig. 1a), and its standard deviation is on the same order (Supplementary Fig. 1b). Such a wind stress curl leads to pronounced upwelling in the California Current system in the Pacific[47] and off Northwest Africa in the Atlantic[48]. In contrast, the wind stress curl during pre-genesis TCs is greater by one order or even two orders of magnitude (Supplementary Fig. 2), which is strong enough to drive surface-subsurface interactions and produce changes in SST.

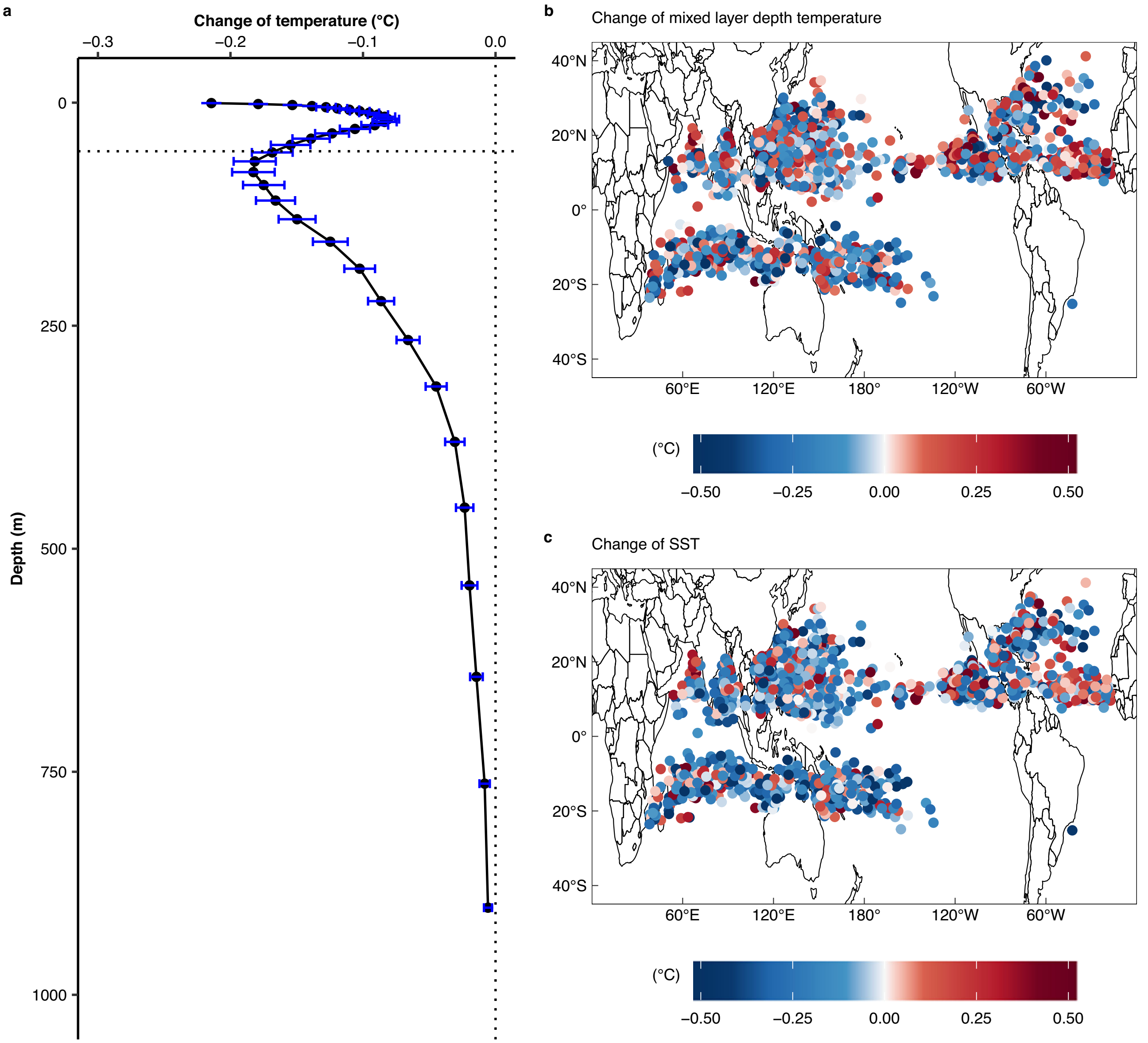

**Fig. 4 | Changes in ocean column temperatures due to pre-genesis tropical cyclones (TCs). a** Composite profiles of ocean temperature changes. The black points denote the mean ocean temperature changes. The blue error bars denote the corresponding standard error of the mean (SEM). Horizontal dotted line denotes the composite mean depth of the 26 °C isotherm (D26). **b** Same as Fig. 2a but for the temperature anomalies at the mixed layer depth. **c** Same as Fig. 2a but for the sea surface temperature (SST) anomalies.

Quantitatively, according to the classical Ekman theory, the Ekman depth is given by $D_E = \sqrt{(2A_z/f)}$,where $A_z$ is the vertical momentum diffusivity and $f$ is the Coriolis parameter. There are few observations for $A_z$; hence, the Ekman depth is typically estimated by empirical formulas. The Ekman depths estimated with four widely applied empirical formulas (Supplementary Table 1) and D26 under each pre-genesis TC are shown in Fig. 3a. Before the TC genesis, D26 is mostly shallower than 121 m (black dots in Fig. 3a). In the deep tropics (5–10° latitude), the mean D26 is approximately 69 m. It increases towards the mid-latitudes (20–30° latitude), which is consistent with the spatial distribution of climatological D26. In the four empirical formulas, $D_E$ is inversely proportional to sin $\phi$ or $\sqrt{\sin\phi}$, where $\phi$ is the latitude. Thus, the estimated $D_E$ increases with the decrease in latitude. In the deep tropics, all estimated values of $D_E$ are much deeper than those of D26. Beyond 10° of latitude around the equator, the estimated $D_E$ are still deeper than D26. The ratios of D26 to $D_E$ are mostly greater than 2 (Fig. 3b) regardless of the empirical formula used. Therefore, we show that the winds and the associated wind stress curl during TC pre-genesis, which were traditionally assumed to be weak, are in fact strong enough to significantly displace the local D26 in the ocean.

## Feedback of subsurface variabilities onto TC genesis

Subsurface ocean perturbations (e.g., D26 variations) due to wind stress and wind stress curl during TC pre-genesis can significantly modify ocean temperatures, including local SSTs, which are essential for consequent TC growth and intensification. Similar to the definition of ΔD26, the differences between temperatures on Day 0 and the undisturbed temperatures (i.e., the median temperatures from Days −10 to −4) denote the changes during TC pre-genesis (Methods). The pre-genesis TCs modify the ocean potential temperature profiles down to approximately 200 m. Even at greater depths, their influence is still noticeable, though weaker (Fig. 4a and Supplementary Fig. 3). Moreover, TC-induced temperature changes reach deeper in the context of deep climatological D26 (Supplementary Fig. 4). The mixed layer is characterized by nearly uniform temperature, salinity, and density. The mixed layer depth is defined as the depth where the temperature is cooler than the temperature at 10 m depth by 0.2°C, following ref. 49. The ocean temperature at the mixed layer depth determines the heat exchanges between the subsurface ocean and the upper ocean mixed layer. Therefore, their changes during TC pre-genesis (Fig. 4b) can have a direct impact on the ocean heat content in the upper ocean. As a result, SSTs are modified significantly (Fig. 4c). The statistically significant dependence of SST variabilities on D26 changes is shown in Fig. 5a. An increase in D26 leads to ocean warming below the surface due to Ekman pumping, while a decrease in D26 reduces ocean temperatures due to the upwelling of cool waters, and wind stresses are strong enough to entrain cold water into the mixed layer from the ocean interior even during TC pre-genesis. Meanwhile, in addition to the changes in D26, the propensity for SST changes is also subject to the background state of D26 (colour shading in Fig. 1b). When the local climatological D26 is shallow, the changes of SSTs are more sensitive to changes in D26. (Supplementary Fig. 5a). When the climatological D26 is deep, the warm water layer in the upper ocean is thick, and SSTs are not easily perturbed. Thus, the SSTs are typically less sensitive to D26 changes during the pre-genesis (Supplementary Fig. 5b). Therefore, both local D26 changes and climatological D26 are important to SST changes.

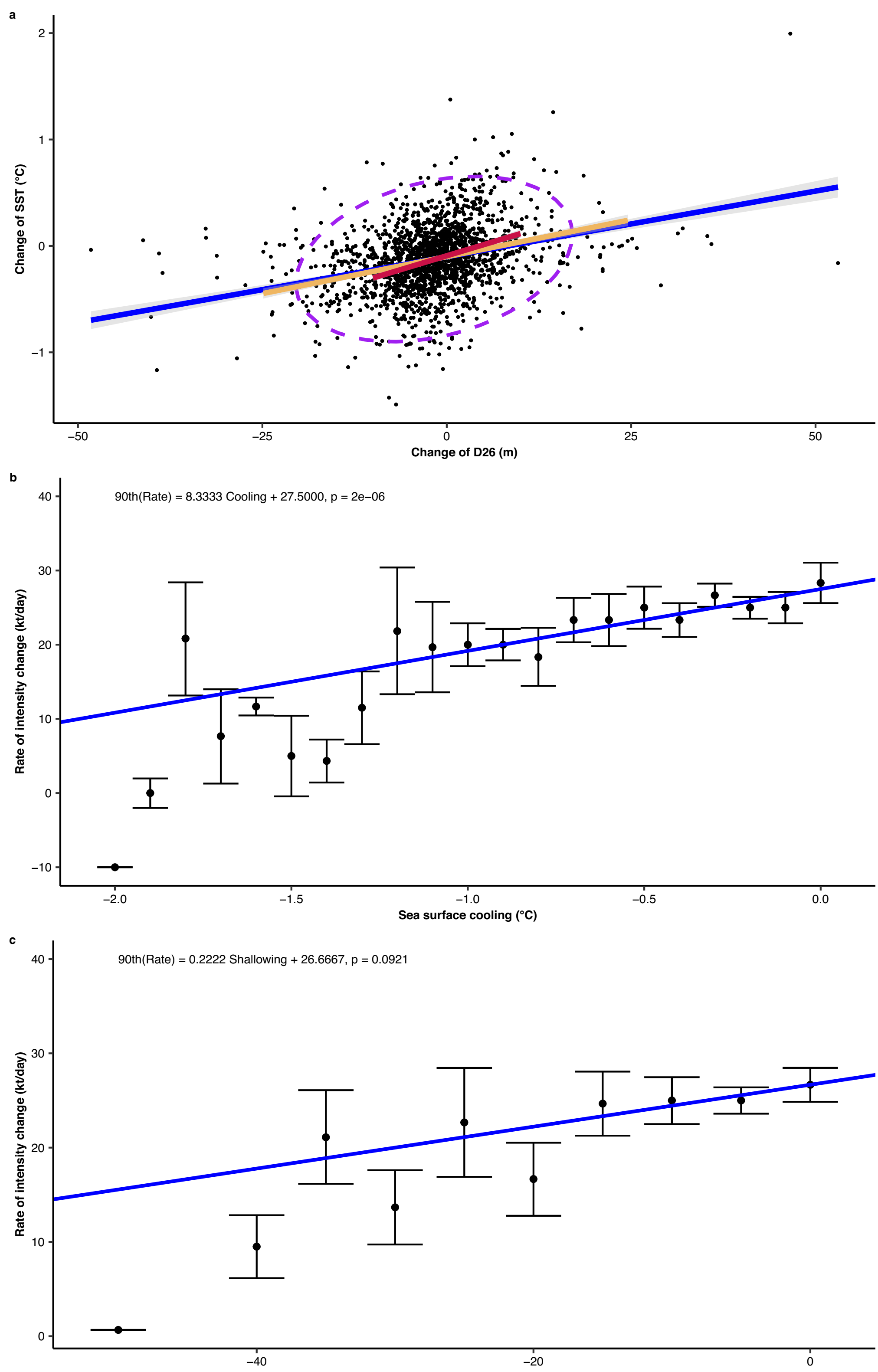

**Fig. 5 | Significant impacts of the subsurface ocean on the genesis of tropical cyclones (TCs). a** Relationship between the changes in the depth of the 26 °C isotherm (D26) and the changes in sea surface temperatures (SSTs). The black line represents the linear regression between the two changes. Grey shading represents the 95% confidence interval. Yellow line is for the linear regression with the range of D26 change from −25 m to 25 m. Red line is for the linear regression with the range of D26 change from −10 m to 10 m. The purple dashed ellipse encloses 95% of all points. **b** Relationship between the changes in SSTs and the rate of TC intensity changes. Black dots represent the 90th percentiles of the TC intensity change rate. The black error bars are the standard errors of the 90th percentile estimated by bootstrapping with 10,000 replicates (Methods). Solid blue line represents the 90th quantile regression, and the regression equation with the p-value is shown on the upper side of the panel. **c** Same as **b** but for relationship between the changes in D26 and the rate of TC intensity change.

For mature TCs, sea surface cooling is known to have a significant negative feedback effect on TC intensification[7,24,25,27,28,50]. A similar conclusion is found for pre-genesis TCs. As shown in Fig. 5b, there is a statistically significant positive correlation between SST changes and the 90th percentile of the TC intensity change rate from Day 0 to Day 3 (a proxy for quantifying TC genesis). The changes in SSTs during TC pre-genesis are determined using daily satellite data[51], Argo data[52], and hourly drifter data[53] (see Methods). The sensitivity of TC genesis to SST changes is 8 knots day$^{-1}$ °C$^{-1}$ (Methods). Therefore, the SST changes due to D26 variations during TC pre-genesis have important influences on the increase in TC intensity. Furthermore, the role of D26 in TC genesis is also supported by the direct relationship between the changes in D26 and the rate of TC intensity change (Fig. 5c).

## Discussion

The key atmospheric properties for TC genesis have long been identified[22], although there has been no general theory for TC genesis[54]. For the ocean, SSTs are found to be important for the entire TC life cycle since they adjust the air-sea heat fluxes. In this study, we show that the subsurface ocean also plays a significant role in TC genesis. The wind speeds and wind stress curls associated with embryonic TCs are much greater than those associated with background normal winds. Thus, they are strong enough to pull up deeper isopycnic layers up to D26. As a result, the ocean temperature profiles can be significantly modified, leading to either warm subsurface anomalies due to surface convergence and heat buildup[40] or cold subsurface anomalies due divergence along with Ekman suction[41]. SST anomalies are produced either by the convergence or divergence and detrainment or entrainment, which then feeds back to the budding TCs. Here, we highlight that TC's entrainment mixing, as well as Ekman upwelling, can be significant during the genesis stage, a notion largely overlooked in existing literature. We demonstrate that, contrary to previous perceptions, TC winds are not weak even during the genesis stage, and TC-subsurface interactions are already occurring at this early stage. Besides, previous studies[55,56] have demonstrated that the TC-induced near-inertial current corresponding to upwelling and downwelling are evident in the recovery stage (after Day 0).

Our study does not deny the effects of other environmental factors (such as vertical wind shear) on TC genesis. The TC genesis issue is complex, and there are many contributors. Genesis is the joint contribution from all factors, but no single factor is the sole factor. D26 can also represent the ocean stratification to some extent, which is determined by both temperature and salinity. Given compelling evidence that salinity can influence TC intensification[57–59], including its ability to suppress cooling by up to 0.3°C[57], the role of salinity in TC genesis warrants greater attention and further investigation. For example, in the eastern North Pacific, the thermocline and D26 are relatively shallow, providing less energy for TC genesis. However, the sea surface salinity is low, resulting in strong salinity-driven ocean stratification, which reduces cooling during the TC genesis stage and potentially promotes TC genesis.

The impacts of the subsurface ocean on TC genesis are captured in state-of-the-art coupled climate models. The High Resolution Model Intercomparison Project (HighResMIP) models[60] permit TCs. Although the simulated TCs in HighResMIP are weaker than the observed TCs, they show an improvement in reproducing the observed TC frequency[61]. The role of the subsurface ocean in TC genesis is estimated with outputs from a 1573-year hierarchy of HighResMIP model simulations (Supplementary Table 2). A deeper D26 indicates a greater tropical cyclogenesis number for both the Northern Hemisphere and the Southern Hemisphere (Supplementary Fig. 6 and Supplementary Fig. 7). It is still challenging to replicate the observed D26 trends in HighResMIP coupled models (Supplementary Fig. 8 and Supplementary Table 3). The bias in D26 might persist in the future scenarios, which potentially leads to a bias in TC genesis projection. This also underscores the need to capture the wind responses to continued warming and the associated vertical structures in the ocean to be able to make reliable projections of TCs in the future.

## Methods

### TC data

The 6-hourly TC best-track datasets are the revised Hurricane Database (HURDAT2)[62] managed by the U.S. National Hurricane Center (NHC) for the North Atlantic, Northeast and North Central Pacific and the best-track archive provided by the Joint Typhoon Warning Center (JTWC) for the other basins. Both NHC and JTWC report TC intensities in terms of the 1-minute mean wind speeds.

We use the H*Wind analyses[63] produced by the Hurricane Research Division of the U.S. National Oceanic and Atmospheric Administration (NOAA) to derive the surface wind stress curl for pre-genesis TCs. The H*Wind software produces a gridded analysis by interpolating and smoothing all available wind speed observations. H*Wind analyses provide more than 1500 snapshots of the TC surface wind field, mainly from 1995 to 2013, over the North Atlantic and North Pacific. The horizontal resolution of these snapshots is 6 km.

### SST data

We use the daily Remote Sensing Systems (REMSS) SST data for 25 years from 1998 to 2022. The spatial resolution is 0.25° × 0.25°. The product is generated by microwave radiometer measurements and an optimum interpolation method[64], which is designed to depict the SSTs at a depth of approximately 1 m without the diurnal warming. Thermal infrared SST measurements have high spatial resolution (~9 km) but only work in cloud-free atmospheres[65]. Passive microwave SST measurements have relatively low spatial resolution (~25 km) but are not affected by clouds or other atmospheric effects[51]. Due to the dense cirrus clouds covering TCs, microwave SSTs have been widely used in TC-related studies[7,25,28,50,58,66–71]. In December 1997, the Tropical Rainfall Measuring Mission Microwave Imager took measurements at 10.7 GHz, which provided the first access to high-quality microwave SST data[51]. Hourly SST data[53] are obtained from surface drifting buoys deployed by the Global Drifter Program (GDP) funded by NOAA. Ocean temperatures (including SSTs) are adopted from Argo profiles maintained by the international Argo program[52]. Argo profiles are identified and downloaded using an R package called argoFloats[72]. We compare the SSTs from drifter, Argo and microwave (Supplementary Fig. 9) and D26 from Argo and GLORYS12v1 (Supplementary Fig. 10), and they are highly consistent.

### Daily ocean temperature data

Daily and global ocean environmental variables are obtained from the GLORYS12v1 product[36]. The product is eddy-resolving with a horizontal resolution of 1/12°. There are 50 vertical levels. The GLORYS12v1 output is largely based on the current real-time global forecasting Copernicus Marine Environmental Monitoring Service (CMEMS) system. GLORYS12v1 is generated by the Nucleus for European Modeling of the Ocean (NEMO) ocean model, which is forced at the surface by the European Center for Medium-Range Weather Forecasts (ECMWF) ERA-Interim and then the ERA5 reanalysis for recent years. In addition, GLORYS12v1 jointly assimilates along-track altimeter data, satellite-based SST, and in-situ temperature and salinity vertical profiles.

### Monthly ocean temperature data

Ocean Reanalysis System 5 (ORAS5) reanalysis monthly ocean temperatures with a horizontal resolution of 1/4° are obtained from ECMWF[73]. The EN4.2.2 analysis monthly ocean temperatures with a horizontal resolution of 1° are obtained from the Met Office Hadley Centre[74]. The Institute of Atmospheric Physics version 4 (IAPv4) global ocean temperature gridded product at 1° horizontal resolution are obtained from the IAP[75].

### Daily near-surface wind data

Daily near-surface winds are obtained from NCEP-National Center for Atmospheric Research (NCAR) Reanalysis 1[76] with a horizontal resolution of 2.5° × 2.5°.

### Definition of pre-genesis TCs

The Dvorak technique based on cloud patterns in satellite imagery is the primary method and "gold standard" for estimating TC intensities[35]. The Dvorak technique quantifies the TC intensity on a scale of 1–8 (at 0.5 intervals) called T-Numbers. For example, T2.0 corresponds to 30 knots, and T2.5 corresponds to 35 knots. The Dvorak technique yields estimates that are internally consistent but biased depending on the intensity[77]. However, Dvorak technique-based estimates have almost zero bias and are as accurate as in-situ or airborne observations when the TC intensity is no more than 35 knots[77]. Thirty-five knots are typically used as the threshold for TC genesis. The closest to and lower than the 35 knots threshold is 30 knots in the 6-hourly best-track datasets. Thus, the records that reach 30 knots for the first time are defined as pre-genesis TCs. Applying the criteria above, we identify 2209 pre-genesis TCs from 1998 to 2022. We use 2032 pre-genesis TCs after removing 177 cases since their genesis locations are less than 25 km away from land.

### Quantile regression and bootstrapping

Quantile regression models are used to estimate the percentiles of the dependent variables, conditional on the values of the explanatory variables. Unlike linear regression models, estimating the standard errors of the coefficients or the p-values of a quantile regression model is difficult mathematically. Bootstrapping provides a good estimate of the standard error of any statistic by approximating the shape of the sampling distribution[78]. In this study, bootstrapping with 10,000 replicates is used to estimate the p-values of quantile regressions.

### TCs in the HighResMIP models

Based on 6-hourly model outputs, a tracking method named "TRACK"[79] is used to identify simulated TCs in the hierarchy of HighResMIP models (Supplementary Table 2). No tuning of the detection parameters is used in the process of tracking (see ref. 61 for details). Here, we use a hierarchy of HighResMIP models based on the Hadley Centre Global Environment Model in the Global Coupled configuration 3.1 (HadGEM3-GC3.1)[80].

## Data availability

All datasets used in this study are publicly available. HURDAT2 data are available from the U.S. National Hurricane Center (https://www.nhc.noaa.gov/data/); JTWC data are available from the Joint Typhoon Warning Center (https://www.metoc.navy.mil/jtwc/jtwc.html?best-tracks); H*Wind analyses are available from Risk Management Solutions (https://www.rms.com/event-response/hwind/legacy-archive/storms); REMSS microwave SST data are available from Remote Sensing Systems (https://data.remss.com/SST/daily/mw/v05.1/); Hourly drifter SST data are available from the Atlantic Oceanographic and Meteorological Laboratory (https://www.aoml.noaa.gov/phod/gdp/hourly_data.php); Argo profile data are available from the Ifremer (https://erddap.ifremer.fr/erddap/tabledap/ArgoFloats.html); ORAS5 data are available from the Copernicus Climate Data Store (https://cds.climate.copernicus.eu/cdsapp#!/dataset/reanalysis-oras5?tab=overview); EN4.2.2 data are available from the Met Office Hadley Centre (https://www.metoffice.gov.uk/hadobs/en4/download-en4-2-2.html); IAPv4 data are available from IAP; NCEP-NCAR Reanalysis 1 data are available from the Physical Sciences Laboratory (https://psl.noaa.gov/data/gridded/data.ncep.reanalysis.html); HighResMIP model outputs are available from the Earth System Grid Federation (ESGF; https://esgf-index1.ceda.ac.uk/search/cmip6-ceda/); and GLORYS12v1 data are available from the Copernicus Marine Service (https://data.marine.copernicus.eu/product/GLOBAL_MULTIYEAR_PHY_001_030/description).

## Code availability

The R package argoFloat is available from the Comprehensive R Archive Network (https://cran.r-project.org/web/packages/argoFloats/). The scripts for generating the figures are available at https://doi.org/10.5281/zenodo.14642370[81].

## References

1. Emanuel, K. Increasing destructiveness of tropical cyclones over the past 30 years. *Nature* **436**, 686–688 (2005).
2. Knight, D. B. & Davis, R. E. Climatology of tropical cyclone rainfall in the southeastern United States. *Phys. Geogr.* **28**, 126–147 (2007).
3. Resio, D. T. & Irish, J. L. Tropical cyclone storm surge risk. *Curr. Clim. Change Rep.* **1**, 74–84 (2015).
4. Woodruff, J. D., Irish, J. L. & Camargo, S. J. Coastal flooding by tropical cyclones and sea-level rise. *Nature* **504**, 44–52 (2013).
5. Pielke, R. A. Jr et al. Normalized hurricane damage in the United States: 1900–2005. *Nat. Hazard. Rev.* **9**, 29–42 (2008).
6. Lin, I. I. et al. A tale of two rapidly intensifying supertyphoons: Hagibis (2019) and Haiyan (2013). *Bull. Am. Meteorol. Soc.* **102**, E1645–E1664 (2021).
7. Lin, I.-I., Wu, C.-C., Pun, I.-F. & Ko, D.-S. Upper-ocean thermal structure and the western North Pacific Category 5 typhoons. Part I: ocean features and the Category 5 typhoons' intensification. *Mon. Weather Rev.* **136**, 3288–3306 (2008).
8. Douris, J. & Kim, G. *WMO Atlas of Mortality and Economic Losses from Weather, Climate and Water Extremes (1970–2019)* (World Meteorological Organization, 2021).
9. Sajjad, M. & Chan, J. C. L. Risk assessment for the sustainability of coastal communities: a preliminary study. *Sci. Total Environ.* **671**, 339–350 (2019).

10. Sharmila, S. & Walsh, K. J. E. Recent poleward shift of tropical cyclone formation linked to Hadley cell expansion. *Nat. Clim. Change* **8**, 730–736 (2018).
11. Daloz, A. S. & Camargo, S. J. Is the poleward migration of tropical cyclone maximum intensity associated with a poleward migration of tropical cyclone genesis? *Clim. Dyn.* **50**, 705–715 (2018).
12. Kossin, J. P., Emanuel, K. A. & Vecchi, G. A. The poleward migration of the location of tropical cyclone maximum intensity. *Nature* **509**, 349–352 (2014).
13. DeMaria, M., Knaff, J. A. & Connell, B. H. A tropical cyclone genesis parameter for the tropical Atlantic. *Weather Forecast.* **16**, 219–233 (2001).
14. Latos, B. et al. The role of tropical waves in the genesis of Tropical Cyclone Seroja in the Maritime Continent. *Nat. Commun.* **14**, 856 (2023).
15. Chang, C. P., Liu, C.-H. & Kuo, H.-C. Typhoon Vamei: an equatorial tropical cyclone formation. *Geophys. Res. Lett.* **30**, 1150 (2003).
16. Chen, S. S., Knaff, J. A. & Marks, F. D. Effects of vertical wind shear and storm motion on tropical cyclone rainfall asymmetries deduced from TRMM. *Mon. Weather Rev.* **134**, 3190–3208 (2006).
17. Teng, H.-F., Kuo, Y.-H. & Done, J. M. Importance of midlevel moisture for tropical cyclone formation in easterly and monsoon environments over the western North Pacific. *Mon. Weather Rev.* **149**, 2449–2469 (2021).
18. Montgomery, M. T. & Farrell, B. F. Tropical cyclone formation. *J. Atmos. Sci.* **50**, 285–310 (1993).
19. Emanuel, K. A. The theory of hurricanes. *Annu. Rev. Fluid Mech.* **23**, 179–196 (1991).
20. Emanuel, K. A. An air-sea interaction theory for tropical cyclones. Part I: steady-state maintenance. *J. Atmos. Sci.* **43**, 585–605 (1986).
21. Leipper, D. F. & Volgenau, D. Hurricane heat potential of the Gulf of Mexico. *J. Phys. Oceanogr.* **2**, 218–224 (1972).
22. Gray, W. M. Hurricanes: their formation, structure and likely role in the tropical circulation. In Meteorology over the Tropical Oceans (ed. Shaw, D. B.) 155–218 (Royal Meteorological Society, 1979).
23. Emanuel, K. 100 years of progress in tropical cyclone research. *Meteorol. Monogr.* **59**, 15.1–15.68 (2018).
24. Emanuel, K. A. Thermodynamic control of hurricane intensity. *Nature* **401**, 665–669 (1999).
25. Lloyd, I. D. & Vecchi, G. A. Observational evidence for oceanic controls on hurricane intensity. *J. Clim.* **24**, 1138–1153 (2011).
26. Jin, F. F., Boucharel, J. & Lin, I. I. Eastern Pacific tropical cyclones intensified by El Niño delivery of subsurface ocean heat. *Nature* **516**, 82–85 (2014).
27. Huang, P., Lin, I. I., Chou, C. & Huang, R.-H. Change in ocean subsurface environment to suppress tropical cyclone intensification under global warming. *Nat. Commun.* **6**, 7188 (2015).
28. Lin, I. I. et al. An ocean coupling potential intensity index for tropical cyclones. *Geophys. Res. Lett.* **40**, 1878–1882 (2013).
29. Palmén, E. On the formation and structure of tropical hurricanes. *Geophysica* **3**, 26–38 (1948).
30. Shay, L. K. & Chang, S. W. Free surface effects on the near-inertial ocean current response to a hurricane: a revisit. *J. Phys. Oceanogr.* **27**, 23–39 (1997).
31. Shay, L. K. *Observations of Inertio-Gravity Waves in the Wake of Hurricane Frederic* Master's thesis, Naval Postgraduate School, (1983).
32. Lin, I. et al. New evidence for enhanced ocean primary production triggered by tropical cyclone. *Geophys. Res. Lett.* **30**, 1718 (2003).
33. Gao, C., Zhou, L., Wang, C., Lin, I. I. & Murtugudde, R. Unexpected limitation of tropical cyclone genesis by subsurface tropical central-north Pacific during El Niño. *Nat. Commun.* **13**, 7746 (2022).
34. Emanuel, K. Tropical cyclones. *Annu. Rev. Earth Planet. Sci.* **31**, 75–104 (2003).
35. Velden, C. et al. The Dvorak tropical cyclone intensity estimation technique: a satellite-based method that has endured for over 30 years. *Bull. Am. Meteorol. Soc.* **87**, 1195–1210 (2006).
36. Jean-Michel, L. et al. The Copernicus global 1/12° oceanic and sea ice GLORYS12 reanalysis. *Front. Earth Sci.* **9**, 698876 (2021).
37. Pun, I.-F., Lin, I. I. & Lo, M.-H. Recent increase in high tropical cyclone heat potential area in the western North Pacific Ocean. *Geophys. Res. Lett.* **40**, 4680–4684 (2013).
38. Lin, I.-I. & Chan, J. C. L. Recent decrease in typhoon destructive potential and global warming implications. *Nat. Commun.* **6**, 7182 (2015).
39. Price, J. F. Upper ocean response to a hurricane. *J. Phys. Oceanogr.* **11**, 153–175 (1981).
40. Emanuel, K. Contribution of tropical cyclones to meridional heat transport by the oceans. *J. Geophys. Res.: Atmos.* **106**, 14771–14781 (2001).
41. Zhang, H., He, H., Zhang, W.-Z. & Tian, D. Upper ocean response to tropical cyclones: a review. *Geosci. Lett.* **8**, 1–12 (2021).
42. Zhang, H. Modulation of upper ocean vertical temperature structure and heat content by a fast moving tropical cyclone. *J. Phys. Oceanogr.* **53**, 493–508 (2023).
43. Shay, L. K., Elsberry, R. L. & Black, P. G. Vertical structure of the ocean current response to a hurricane. *J. Phys. Oceanogr.* **19**, 649–669 (1989).
44. Vallis, G. K. *Atmospheric and Oceanic Fluid Dynamics* (Cambridge University Press, 2017).
45. Castelao, R. M. & Barth, J. A. Upwelling around Cabo Frio, Brazil: the importance of wind stress curl. *Geophys. Res. Lett.* **33**, L03602 (2006).
46. Bakun, A. & Nelson, C. S. The seasonal cycle of wind-stress curl in subtropical eastern boundary current regions. *J. Phys. Oceanogr.* **21**, 1815–1834 (1991).
47. Huyer, A. Coastal upwelling in the California current system. *Prog. Oceanogr.* **12**, 259–284 (1983).
48. Mittelstaedt, E. The upwelling area off Northwest Africa—a description of phenomena related to coastal upwelling. *Prog. Oceanogr.* **12**, 307–331 (1983).
49. de Boyer Montégut, C., Madec, G., Fischer, A. S., Lazar, A. & Iudicone, D. Mixed layer depth over the global ocean: an examination of profile data and a profile-based climatology. *J. Geophys. Res.: Oceans* **109**, C12003 (2004).
50. Lin, I.-I., Pun, I.-F. & Wu, C.-C. Upper-ocean thermal structure and the western North Pacific Category 5 typhoons. Part II: dependence on translation speed. *Mon. Weather Rev.* **137**, 3744–3757 (2009).
51. Wentz, F. J., Gentemann, C., Smith, D. & Chelton, D. Satellite measurements of sea surface temperature through clouds. *Science* **288**, 847–850 (2000).
52. Wong, A. P. S. et al. Argo data 1999–2019: two million temperature-salinity profiles and subsurface velocity observations from a global array of profiling floats. *Front. Mar. Sci.* **7**, 700 (2020).
53. Elipot, S., Sykulski, A., Lumpkin, R., Centurioni, L. & Pazos, M. A dataset of hourly sea surface temperature from drifting buoys. *Sci. Data* **9**, 567 (2022).
54. Sobel, A. H. et al. Tropical cyclone frequency. *Earths Future* **9**, e2021EF002275 (2021).
55. Zhang, H. et al. Upper ocean response to typhoon Kalmaegi (2014). *J. Geophys. Res.: Oceans* **121**, 6520–6535 (2016).
56. Zhang, H. et al. Net modulation of upper ocean thermal structure by Typhoon Kalmaegi (2014). *J. Geophys. Res.: Oceans* **123**, 7154–7171 (2018).
57. Balaguru, K. et al. Ocean barrier layers' effect on tropical cyclone intensification. *Proc. Natl Acad. Sci. USA* **109**, 14343–14347 (2012).
58. Balaguru, K., Foltz, G. R., Leung, L. R. & Emanuel, K. A. Global warming-induced upper-ocean freshening and the intensification of super typhoons. *Nat. Commun.* **7**, 13670 (2016).

59. Balaguru, K. et al. Pronounced impact of salinity on rapidly intensifying tropical cyclones. *Bull. Am. Meteorol. Soc.* **101**, E1497–E1511 (2020).
60. Haarsma, R. J. et al. High Resolution Model Intercomparison Project (HighResMIP v1.0) for CMIP6. *Geosci. Model Dev.* **9**, 4185–4208 (2016).
61. Roberts, M. J. et al. Projected future changes in tropical cyclones using the CMIP6 HighResMIP multimodel ensemble. *Geophys. Res. Lett.* **47**, e2020GL088662 (2020).
62. Landsea, C. W. & Franklin, J. L. Atlantic hurricane database uncertainty and presentation of a new database format. *Mon. Weather Rev.* **141**, 3576–3592 (2013).
63. Powell, M. D., Houston, S. H., Amat, L. R. & Morisseau-Leroy, N. The HRD real-time hurricane wind analysis system. *J. Wind Eng. Ind. Aerodyn.* **77**, 53–64 (1998).
64. Reynolds, R. W. & Smith, T. M. Improved global sea surface temperature analyses using optimum interpolation. *J. Clim.* **7**, 929–948 (1994).
65. Deschamps, P. Y. & Phulpin, T. Atmospheric correction of infrared measurements of sea surface temperature using channels at 3.7, 11 and 12 μm. *Boundary Layer Meteorol.* **18**, 131–143 (1980).
66. Yang, Y. J. et al. The role of enhanced velocity shears in rapid ocean cooling during Super Typhoon Nepartak 2016. *Nat. Commun.* **10**, 1627 (2019).
67. Pun, I.-F., Knaff, J. A. & Sampson, C. R. Uncertainty of tropical cyclone wind radii on sea surface temperature cooling. *J. Geophys. Res.: Atmos.* **126**, e2021JD034857 (2021).
68. Pun, I.-F., Lin, I. I., Lien, C.-C. & Wu, C.-C. Influence of the size of Supertyphoon Megi (2010) on SST cooling. *Mon. Weather Rev.* **146**, 661–677 (2018).
69. Lin, I. I. et al. The interaction of Supertyphoon Maemi (2003) with a warm ocean eddy. *Mon. Weather Rev.* **133**, 2635–2649 (2005).
70. Xu, J., Wang, Y. & Tan, Z.-M. The relationship between sea surface temperature and maximum intensification rate of tropical cyclones in the North Atlantic. *J. Atmos. Sci.* **73**, 4979–4988 (2016).
71. Vincent, E. M. et al. Processes setting the characteristics of sea surface cooling induced by tropical cyclones. *J. Geophys. Res.: Oceans* **117**, C02020 (2012).
72. Kelley, D. E., Harbin, J. & Richards, C. argoFloats: an R package for analyzing argo data. *Front. Mar. Sci.* **8**, 635922 (2021).
73. Zuo, H., Balmaseda, M. A., Tietsche, S., Mogensen, K. & Mayer, M. The ECMWF operational ensemble reanalysis-analysis system for ocean and sea ice: a description of the system and assessment. *Ocean Sci.* **15**, 779–808 (2019).
74. Good, S. A., Martin, M. J. & Rayner, N. A. EN4: quality controlled ocean temperature and salinity profiles and monthly objective analyses with uncertainty estimates. *J. Geophys. Res.: Oceans* **118**, 6704–6716 (2013).
75. Cheng, L. et al. IAPv4 ocean temperature and ocean heat content gridded dataset. *Earth Syst. Sci. Data* **2024**, 1–56 (2024).
76. Kalnay, E. et al. The NCEP/NCAR 40-year reanalysis project. *Bull. Am. Meteorol. Soc.* **77**, 437–472 (1996).
77. Knaff, J. A., Brown, D. P., Courtney, J., Gallina, G. M. & Beven, J. L. An evaluation of Dvorak technique-based tropical cyclone intensity estimates. *Weather Forecast.* **25**, 1362–1379 (2010).
78. Efron, B. Bootstrap methods: another look at the jackknife. *Ann. Stat.* **7**, 1–26 (1979).
79. Hodges, K., Cobb, A. & Vidale, P. L. How well are tropical cyclones represented in reanalysis datasets? *J. Clim.* **30**, 5243–5264 (2017).
80. Roberts, M. J. et al. Description of the resolution hierarchy of the global coupled HadGEM3-GC3.1 model as used in CMIP6 HighResMIP experiments. *Geosci. Model Dev.* **12**, 4999–5028 (2019).
81. Gao, C. Rendering scripts and source data for figures in 'Crucial role of subsurface ocean variability in tropical cyclone genesis'. Zenodo https://doi.org/10.5281/zenodo.14642370 (2025).

## Acknowledgements
We thank Philip Klotzbach and Kerry Emanuel for valuable discussions. We acknowledge the support by the National Natural Science Foundation of China (Grant nos. 42125601 and 421925648), the National Key Research and Development Program of China (Grant nos. 2023YFF0805300 and 2019YFA0606701), the Strategic Priority Research Program of the Chinese Academy of Sciences (Grant no. XDB42000000), and the Development fund of South China Sea Institute of Oceanology of the Chinese Academy of Sciences (Grant no. SCSIO202208).

## Author contributions
C.G., L.Z., I.I.L., and C.W. conceived the study; C.G. performed the analysis and drew all the figures; C.G. and L.Z. wrote the initial manuscript. C.G., L.Z., I.I.L., C.W., S.G., F.-F.J., and R.M. discussed and approved the results and conclusions of this article.

## Competing interests
The authors declare no competing interests.

## Additional information
**Supplementary information** The online version contains supplementary material available at https://doi.org/10.1038/s41467-025-56433-5.

**Correspondence** and requests for materials should be addressed to Lei Zhou.

**Peer review information** *Nature Communications* thanks the anonymous reviewers for their contribution to the peer review of this work. A peer review file is available.